
\tolerance=10000
\input phyzzx

\font\mybb=msbm10 at 12pt
\def\bbbb#1{\hbox{\mybb#1}}
\def\Z {\bbbb{Z}}
\def\R {\bbbb{R}}

\REF \ShSh{J. Scherk and J.H. Schwarz,     {  Nucl. Phys.}~{\bf B153} (1979) 
61.}
\REF\bergy{E. Bergshoeff, M. de Roo, M.B. Green, G. Papadopoulos and P.K. 
Townsend,
Nucl. phys. {\bf B470} (1996) 113.}
\REF\fibs{ I.V.~Lavrinenko, H.~Lu and C.N.~Pope, 
Class.\ Quant.\ Grav.\  {\bf 15}, 2239 (1998),  hep-th/9710243.}
\REF\kuri{ N.~Kaloper, R.R.~Khuri and R.C.~Myers, 
Phys.\ Lett.\ B {\bf 428}, 297 (1998),
                {  hep-th/9803006}.}
\REF\gage{ E. Bergshoeff ,  M. de Roo ,  E. Eyras, Phys.Lett. B413 (1997) 70-78;
hep-th/9707130.}
\REF\wall{ P.M. Cowdall,  H. Lu,  C.N. Pope,  K.S. Stelle, P.K. Townsend, 
Nucl.Phys.
B486 (1997) 49; hep-th/9608173 .}
\REF\myers{ N. Kaloper and R. Myers, The
O(dd) story of massive supergravity, JHEP {\bf 05} (1999) 010, 
hepth{9901045}. }
\REF\mf{C.M. Hull,  Massive string theories from M-theory and F-theory,  JHEP 
{\bf
9811} (1998) 027, hepth{9811021}.}
\REF\ort{ P.~Meessen and T.~Ortin,  Nucl. Phys. {\bf B541}{(1999)}{195} { 
hep-th/9806120}.}
\REF\cow{P.M. Cowdall, hep-th/0009016.}
\REF\mee{
J.~Gheerardyn and P.~Meessen,
Phys.\ Lett.\ B {\bf 525}, 322 (2002)
 hep-th/0111130.}
\REF\atish{A. Dabolkhar and C.M. Hull, in preparation.}
\REF\CJ{E. Cremmer and B. Julia, Phys. Lett. {\bf 80B} (1978) 48; Nucl.
Phys. {\bf B159} (1979) 141.}
\REF\julia {B. Julia in {\it Supergravity and Superspace}, S.W.
Hawking and M.
Ro\v cek, C.U.P.
Cambridge,  1981. }
\REF\HT{C.M. Hull and P.K. Townsend, Nucl. Phys. {\bf B438} (1995) 109,
hep-th/9410167.}
\REF\bergnew{E. Bergshoeff, U. Gran and D. Roest, in preparation.}
\REF\nct{C.M. Hull, {\it Phys. Rev.} {\bf D30} (1984) 760; C.M. Hull,
{\it Phys. Lett.} {\bf 142B} (1984) 39; C.M. Hull, {\it Phys. Lett.} {\bf
148B} (1984) 297; C.M. Hull, {\it Physica} {\bf 15D} (1985) 230; Nucl.
Phys. {\bf B253} (1985) 650;  C.M. Hull, {\it Class. Quant. Grav.} {\bf
2} (1985) 343.}
\REF\DeWolfeEU{
O.~DeWolfe, T.~Hauer, A.~Iqbal and B.~Zwiebach,
Adv.\ Theor.\ Math.\ Phys.\  {\bf 3}, 1785 (1999); hep-th/9812028.}
\REF\DeWolfePR{
O.~DeWolfe, T.~Hauer, A.~Iqbal and B.~Zwiebach,
Adv.\ Theor.\ Math.\ Phys.\  {\bf 3}, 1835 (1999)
 hep-th/9812209.
}
\REF\GHT{M.B. Green, C.M. Hull and P.K. Townsend,  Phys. Lett. {\bf 382} (1996) 
65,
hep-th/9604119.}




%

\def \aa {\alpha}

\def \dd {\delta}
\def \ee {\epsilon}
\def \ffi {{\phi ^i}}

\def \ll {\lambda}
\def \mm {\mu}

\def \th {\theta}

 \def \ggg {\Gamma}

\def \ti {\tilde}

\def \2 {{1 \over 2}}
\def \3 {{1 \over 3}}
\def \4 {{1 \over 4}}
\def \5 {{1 \over 5}}
\def \6 {{1 \over 6}}
\def \7 {{1 \over 7}}
\def \8 {{1 \over 8}}
\def \9 {{1 \over 9}}
\def \0 { \infty}

\def\++ {{(+)}}
\def \- {{(-)}}
\def\+-{{(\pm)}}

\def\ek {\eqn\abc$$}

\def \qq {\qquad}


 \def\unit{\hbox to 3.3pt{\hskip1.3pt \vrule height 7pt width .4pt \hskip.7pt
\vrule height 7.85pt width .4pt \kern-2.4pt
\hrulefill \kern-3pt
\raise 4pt\hbox{\char'40}}}
\def\II{{\unit}}
\def\cM {{\cal{M}}}

\def\nup#1({Nucl.\ Phys.\  {\bf B#1}\ (}



\Pubnum{ \vbox{  \hbox {QMUL-PH-02-05}  
\hbox{hep-th/0203146}} }
\pubtype{}
\date{March 2002}

\titlepage

\title {\bf  Gauged $D=9$ Supergravities and Scherk-Schwarz Reduction.}

\author{C.M. Hull}
\address{Physics Department, Queen Mary, University of London,
\break Mile End Road, London E1 4NS, U.K.}
\andaddress{Newton Institute, 20 Clarkson Road, Cambridge CB3 0EH, U.K.}

\abstract {
Generalised Scherk-Schwarz reductions in which
compactification on a circle is accompanied by a twist with an element of a 
global
symmetry $G$ 
typically lead to gauged supergravities and
are classified by the monodromy matrices, up to conjugation by 
the global symmetry.
For compactifications of IIB supergravity on a circle, $G=SL(2,\R)$
and there are three distinct gauged supergravities that result, 
corresponding to monodromies in the three conjugacy classes of $SL(2,\R)$.
There is one gauging of the compact $SO(2)$ subgroup of the $SL(2,\R)$
and two distinct gaugings of non-compact $SO(1,1)$ subgroups, embedded 
differently
in $SL(2,\R)$.  
The non-compact gaugings can  be obtained from the compact one via an analytic
continuation of the kind used in $D=4$ gauged supergravities.
For the superstring, the monodromy must be in $SL(2,\Z)$, and the distinct 
theories
correspond to $SL(2,\Z)$ conjugacy classes.
The theories consist of  two infinite classes with quantised mass parameter
$m=1,2,3,...$,    three exceptional theories corresponding to elliptic conjugacy classes, and
a  set of sporadic theories corresponding to hyperbolic conjugacy classes.

 }

\endpage



\chapter{Scherk-Schwarz Reductions and Generalisations}

The Scherk-Schwarz mechanism  and its generalisations
[\ShSh-\mee]
 introduces mass parameters   into toroidal  compactifications of supergravities 
and
string theories. If the original theory in $D'$ dimensions has a global symmetry
$G'$ acting on fields
$ \phi$ by $ \phi\to g( \phi)$, then in a  generalised Scherk-Schwarz reduction 
or
twisted reduction the fields are not independent of the internal coordinates, 
but
are chosen to depend on the torus coordinates $y$ through an ansatz
$$ \phi (x^\mm,y) = g_y(\ffi(x^\mm))
\eqn\ansa$$
 for some $y$-dependent symmetry transformation $g_y=g(y)$ in $ G'$. Usually the
twist will be contained in the subgroup $K'\subseteq G'$ of  symmetries of the 
action
in $D'$ dimensions, but in some cases it is   possible to twist by symmetries 
that
are symmetries of the field equations only (an example of this is given in
[\fibs]); here only twists in $K'$ will be considered.
 In some cases  such a generalised reduction  leads to a spontaneous breaking of
supersymmetry [\ShSh]. Typically,  it results in a gauging of the reduced
theory; see e.g. [\gage,\wall,\ort]. In such cases, if the standard reduction 
without
twists from $D'$  dimensions gives a theory in   $D$ dimensions with duality
symmetry $G$ and symmetry of the action $K\subseteq G$, then twisting with an
element of
$K'\subseteq K$ will result in a gauging of a subgroup $L$ of $K$. 
Consider   compactifications from $D'=D+1$ dimensions to $D$ dimensions
on a circle, with periodic coordinate $y \sim y+1$. For example, for reducing a
theory with a linearly realised $G'=U(1)$ symmetry on a circle, a massless field
$\phi$ of charge
$q$ can be given a $y$ dependence
$\phi(x,y)= e^{2\pi iqmy}\ffi(x)$, so that  the field $\ffi(x)$ is given  a mass 
of
$qm$.

The map $g(y)$ is not periodic, but
 has
 a {\it monodromy}
$$ {\cal M} (g)= g(1)g(0)^{-1}
\eqn\abc$$ for some ${\cal M}$ in $G'$. For maps of the form
$$g(y)= \exp (My)
\eqn\ansatz$$ for some
 Lie algebra element $M$,   the monodromy is
$$ {\cal M} (g)= \exp M
\eqn\mono$$ Then
$$ M=g^{-1}\partial _y g
\eqn\abc$$ is proportional to the mass matrix of the dimensionally reduced 
theory and is
independent of $y$.

The  Lie algebra element $M$ generates a one-dimensional subgroup $L'$ of $G'$, 
and
this group becomes   gauged in the dimensionally reduced theory. For a field 
$\phi$
transforming in some representation of
$G'$ with $M$ acting through some matrix $\bar M$, so that under $G'$, $\phi$
transforms as 
$\dd \phi = \ll \bar M \phi$ (with infinitesimal parameter $\ll$), it is
straightforward to show that on dimensional reduction to $D'-1$ dimensions, the
derivative of $\phi$ becomes the gauge covariant derivative
$D\phi = d \phi +A \bar M \phi$, where $A$ is the 1-form gauge potential arising
from the reduction of the metric (the graviphoton), indicating that $L'$ has 
become
a local symmetry for which the gauge field is the graviphoton. For  Scherk-Schwarz 
reduction on
$T^n$ with the twistings for the $n$ circles generated by $n$
commuting matrices
$M_1,\dots , M_n$, the resulting gauge group is the abelian group
$L'_1\times...\times L'_n$ generated by the $M_i$.

The next question is whether two different choices of $g(y)$ give inequivalent
theories. The   ansatz breaks the symmetry $G'$ down to the subgroup preserving
$g(y)$, consisting of those $h$ in $G'$ such that $h^{-1} g(y) h=g(y)$. Acting 
with
a general constant element  $k$ in $K'$ will change the mass-dependent terms, 
but
will give a
$D-1$ dimensional theory related to the original one via the field refinition
$\phi \to k(\phi)$. This same theory could have been obtained directly via a
 reduction using
$k^{-1} g(y) k$ instead of $g(y)$, so two choices of $g(y)$ in the same 
conjugacy
class give equivalent reductions (related by field-redefinitions). 
A given monodromy can result from infinitely many different mass matrices [\mf], but these
all give physically equivalent results (if all the Kaluza-Klein modes are kept) [\atish].
As a result, 
the
reductions are classified by  conjugacy classes of the monodromy-matrix $\cal M$ [\mf].

The map $g(y)$ is a  local section of a principal fiber  bundle over the circle 
with
fibre $G'$ and monodromy
${\cal M}  (g)$ in $G'$. Such a bundle is constructed from $I\times G'$, where
$I=[0,1]$ is the unit interval, by gluing the ends of the interval together with 
a
twist of the fibres by the monodromy
${\cal M} $. Two such bundles with monodromy in the same $G'$-conjugacy class 
are
equivalent.

Of particular interest are the    supergravity theories in  $D'=D+1$ dimensions 
with
rigid duality symmetry $G'$ and scalars taking values in
$G'/H'$ [\CJ,\julia], which  can be  Scherk-Schwarz-reduced on a circle to $D$
dimensions. The reduction requires the choice of a map $g(y)$ of the form 
\ansatz\
from
$S^1$ to $G'$, which then determines the $y$-dependence of the fields   through 
the
ansatz
\ansa, and any choice of Lie algebra element $M$ is allowed.

In the quantum theory, the symmetry group $G'$ is broken to a discrete sub-group
$G'(\Z)$ [\HT]. A consistent twisted reduction of a string or M-theory, whose
low-energy effective theory is the supergravity theory considered above, then
requires that the monodromy be in the U-duality group $G(\Z)$. (In the classical
supergravity theory, any element of $G$ can be used as the monodromy.) Then
  the choice of $M$ is restricted by the constraint that $e^M$ should be in 
$G'(\Z)$.
As before, if two theories have $M$-matrices $M,\ti M$ related by
  $M=k\ti Mk^{-1}$ where $k$ is in $G'$, the theories are related by field
redefinitions. However, the data needed to specify the quantum theory  includes 
the
charge lattice  $\ggg_p$ giving the allowed values of the quantised
$p$-brane charge, and there is such a lattice for each of the values of 
$p$ arising in the   theory. For each $p$, the $p$-brane charges will transform 
in
some  representation $R_p$ of $G'$ and so the $G'$ transformation $k$   taking 
$M$
to $\ti M$ will take $\ggg_p$ to a new lattice $
\ti \ggg_p$. Thus the theories specified by $(M,\ggg_p)$ and $(\ti M,\ti 
\ggg_p)$ 
are related by field redefinitions and so are physically equivalent. One way to
classify the distinct theories is to fix the lattices $\ggg_p$  and ask which
monodromy matrices give distinct  theories. The subgroup of $G'$ which preserves 
the 
  lattices $\ggg_p$ is the discrete U-duality subgroup, which
  will be denoted $G'(\Z)$, and $G'(\Z)\cap K'$ will be denoted $K'(\Z)$. Then
acting with an element $k$ of
 $K'(\Z)$ will preserve the lattices but change $M$ to $\ti M=k Mk^{-1}$.
 Then the two monodromies
 ${\cal M}$ and $\ti {\cal M}$ will be elements of $K'(\Z)$ in the same 
 $K'(\Z)$ conjugacy class.
 Thus, for   given charge lattices, two theories with $K'(\Z)$  monodromies that 
 are  related by $\ti M=k Mk^{-1}$ for some $k \in K'(\Z)$
 will be physically equivalent, as they are related by a field redefinition.
 Thus the distinct theories correspond to the distinct  
  $K'(\Z)$ conjugacy classes [\mf].

\chapter{Scherk-Schwarz Reduction of IIB Supergavity on $S^1$.}

 The type IIB supergravity theory has $G=SL(2,\R)$ global symmetry and any 
element
$M$ of the
$SL(2,\R)$  Lie algebra
 can be used in the ansatz \ansa,\ansatz\ to give a Scherk-Schwarz reduction to
9-dimensions to obtain a class of massive 9-dimensional supergravity theories. 
Such
reductions for particular elements of $SL(2,\R)$ were given in
[\bergy,\kuri,\fibs],  and the   general class of $SL(2,\R)$ reductions of IIB
supergravity was obtained   in  [\ort,\cow,\mee]. This gives a 3-parameter
family of theories, specified by the choice of matrix 
$$M=\pmatrix{ m_1 & m_2+m_3 \cr m_2-m_3 & -m_1 \cr }\eqn\massm$$ The details of 
the 
reduction of the bosonic sector of the supergravity theory for general $M$ of 
this
form were given in [\ort,\cow]. Note that this ansatz does not allow
the  monodromy
to be an arbitrary
$SL(2,\R)$ group element, but requires it to be in the image of the exponential 
map.
Acting with an $SL(2,\R)$ transformation leaves the mass-independent part of the
theory unchanged but changes the mass matrix  by
$SL(2,\R)$ conjugation, and  any two theories related by such a field 
redefinition
are physically equivalent. There are then three distinct classes of inequivalent
theories, corresponding to the hyperbolic, elliptic and parabolic $SL(2,\R)$
conjugacy classes, represented by monodromy matrices of the form
$${\cal M}_h=\pmatrix{ e^a & 0 \cr 0 & e^{-a} \cr }, \qq {\cal M}_e=\pmatrix{
\cos \theta & \sin \theta \cr - \sin \theta & \cos \theta \cr } , \qq {\cal
M}_p=\pmatrix{ 1 & a \cr 0 & 1 \cr }
\eqn\monc$$ respectively,  generated by the matrices
$${ M}_h=\pmatrix{ a & 0 \cr 0 & -a \cr }, \qq { M}_e=\pmatrix{ 0 &  \theta \cr 
-  
\theta & 0 \cr } , \qq { M}_p=\pmatrix{ 0 & a \cr 0 & 0 \cr }
\eqn\gens$$ and each class is specified by a single coupling constant ($a$ or 
$\th$).
Thus the 3-parameter family of theories splits into three equivalence classes, 
with
all the theories in a given class related by field redefinitions and  rescalings 
of
the single coupling constant.

In $D=10$, the duality group is $K'=SL(2,\R) $
and there  are two 2-form fields $\hat B_2^i$ ($i=1,2$) transforming as an 
$SL(2,\R)$
doublet.
On reduction to 
   $D=9$, the duality group $K$ is $SL(2,\R) \times \R$ and 
the $\hat B_2^i$ reduce to a doublet of 2-forms $  B_2^i$ and a doublet of 
1-forms
$  B_1^i$. In addition, there is a
third vector field $A_1$ from the reduction of the metric, and this is an 
$SL(2,\R)$
singlet.
The 4-form potential in $D=10$ gives a 4-form $C_4$ and a 3-form $C_3$, but the
self-duality constraint in $D=10$ implies that $C_4$ is the dual of $C_3$ in 
$D=9$,
and the theory can be written in terms of $C_3$ alone [\ort,\cow,].
The field strengths
for these gauge fields include
$$\eqalign{
H_2^i & = dB_1^i -M^i{}_j B_2^j
\cr
H_3^i & = dB_2^i -H_2^i \wedge A_1
\cr
G_4&=dC_3 +\2 \ee_{ij}(-B_1^i\wedge H_3^j
+[B_2^i+A_1\wedge B_1^i]\wedge H_2^j)
\cr}
\eqn\abc$$
In particular, these are invariant under the following gauge symmetry with 
1-form parameter $\ll_1^i$
$$\eqalign{
\dd B_1^i&= M^i{}_j \ll_1^j
\cr
\dd B_2^i&= d\ll_1^i
\cr
\dd C_3&=-\2  \ee_{ij} \ll_1^i \wedge H_2^j
\cr}
\eqn\abc$$
This Stuckelberg symmetry is a shift symmetry for $B_1$ and, when $M^i{}_j$ is invertible,
can be used to set $B_1^i=0$,
so that the two 1-forms $B_1^i$ are eaten by the
2-forms $B_2^i$, which become massive.
For the parabolic case in which $M$ is not invertible, one of the 
$B_1^i$ is eaten by one of the $B_2^i$, which becomes massive, so the physical spectrum has
one massive 2-form, a massless 2-form and a massless 1-form gauge field.
The  action includes  the terms 
$$ \int    G_{ij}H_2^i \wedge * H_2^j +  g_{ij}H_3^i \wedge * H_3^j 
\eqn\abc$$
with scalar-dependent matrices $g_{ij}(\phi),G_{ij}(\phi)$ given explicitly in
[\ort,\cow].
For the case in which $M$ is invertible,
this becomes
$$ \int    \ti G_{ij}B_2^i \wedge * B_2^j +  g_{ij}DB_2^i \wedge * DB_2^j 
\eqn\acty$$
 in   the gauge $B_1^i=0$, where
$$  \ti G_{ij} =   G_{kl}M^k{}_i M^l{}_j
\eqn\abc$$
and
$$DB_2^i= dB_2^i +  M^i{}_j B_2^j \wedge A_1
 \eqn\abc$$
is a covariant derivative invariant under the gauge transformation
$$ \dd A_1=d \aa, \qq \dd B_2^i =-\aa M^i{}_j B_2^j
\eqn\gage$$
The first term in \acty\
is a mass term for the $B_2$   field, and \gage\ indicates that the  symmetry
with paramter $\aa(x)$
is the 1-dimensional subgroup of $SL(2,\R)$ generated by $M$, which has been 
gauged,
with gauge field $A_1$; this is confirmed by checking the other sectors of the 
the
theory.
The gauged subgroup in the elliptic case is  the compact
rotation group $SO(2)$ of matrices generated by $M_e$, while in the hyperbolic 
 case it is the non-compact group $SO(1,1)$  generated by
$M_h$.
The parabolic case is similar, and is the gauging of the
of the
non-compact group $SO(1,1)$  generated by $M_p$.
The parameter $a$ or $\th$ is then the gauge coupling constant.
In [\bergnew], it was conjectured that the hyperbolic case corresponds
to an
$SO(1,1)$ gauging.

The group manifold $SL(2,\R)$ with the Cartan-Killing metric is a Lorentzian 
space
with signature $(+,+,-)$. The three distinct theories arise from gauging the
1-dimensional  subgroup of  $SL(2,\R)$  generated by a Lie algebra element that 
is
timelike (the elliptic case), spacelike (the hyperbolic case) or null (the 
parabolic
case). 

The elliptic $SO(2)$ gauging was considered in detail in [\cow]. The other two 
cases
can be obtained from this using the analytic continuation techniques of [\nct].
Consider starting from the
$SO(2)$ gauging with
$$M= g \pmatrix{ 0 &  1 \cr -  
1 & 0 \cr }
\ek
where $g=m_3$.
Consider acting on this theory with the $SL(2,\R)$ transformation
$$k(t)= \pmatrix{ e^{t} &  0 \cr 0 & -e^{-t} \cr }
\ek
for some parameter $t$.
This will take the theory to
a similar theory, but with $M$ replaced by $M'=k(-t)Mk(t)$, so that a conjugate
$SO(2)$ subgroup has been gauged.
This is of course  equivalent 
to the original $SO(2)$ gauging, via a field redefinition.
Next rescale the coupling constant $g \to g e^{-2t}$, so that  
$M$ 
is now is now replaced with
$$M(t)=e^{-2t} k(-t)Mk(t)= g \pmatrix{ 0 &  1 \cr -  
\xi & 0 \cr }
\eqn\sdfhkdf$$
where
$$\xi= e^{-2t}\ek 
throughout the action and supersymmetry transformations.
For all finite real $t$, this gives an $SO(2)$ gauging
equivalent to the original one.
However, taking the limit $t \to \infty$ gives a well-defined theory, but with 
$\xi=0$ in \sdfhkdf, so that the generator is now the parabolic generator
$M_p$ in  \gens\ with $a=g$.
Similarly, continuing to $\xi=-1, t=i\pi/2$, $M(t)$ becomes
the generator of a hyperbolic subgroup, congruent to $M_h$ in \gens\ with $a=g$.
By the arguments of [\nct], these limits of the original $t$-dependent theory 
are
guaranteed to give supersymmetric gauged supergravity theories, giving an 
independent
check that the generalised Scherk-Schwarz reduction gives a locally supersymmetric
theory.

The mass  matrix $M$ \massm\  corresponds to a vector $v=( m_1,m_2,m_3)$ that
transforms as a vector under
$SO(2,1)\sim SL(2,\R)$, and has norm $m_1^2 +m_2^2-m_3^2$
with respect to the Cartan-Killing metric.
The $SO(2)$ gauging has $v=(0,0,m_3)$ and acting with $k(t)$ corresponds to a 
boost
with rapidity $t$ taking $v$ to another timelike vector of the same norm.
The limit $t\to \infty$ is an infinite boost taking $v$ to a null vector
proportional to $(0,1,1)$ and requires a rescaling   of the components of the 
vector.
Continuing to $t=i\pi/2$, the boost becomes a \lq rotation' taking $v$
to a spacelike vector $(0,m,0)$.

\chapter{Scherk-Schwarz Reduction of IIB Superstring on $S^1$.}

In the quantum IIB theory,  the quantization of string and 5-brane charges  
breaks
the classical $SL(2,\R)$ invariance to   the discrete
$SL(2,\Z)
$ U-duality symmetry    [\HT]. The quantum-consistent Scherk-Schwarz reductions 
of
this theory to 9 dimensions are those for which the monodromy is in $SL(2,\Z)$. 
For
given string and 5-brane charge lattices, acting with an $SL(2,\Z)
$ transformation $k$  will preserve the lattices but change  the monodromy $\cM 
\to
k\cM k^{-1}$. This will take the theory to a physically equivalent one, so that 
the
distinct theories  are represented by monodromies in the distinct
$SL(2,\Z)
$ conjugacy classes [\mf]. 
The $SL(2,\Z)
$ conjugacy classes
have been discussed in [\DeWolfeEU,\DeWolfePR].
There is the trivial class $\cM=1$, together with $\cM=-1$. For any conjugacy 
class $\cM$, $-\cM $ and $\pm \cM^{-1}$   also represent
conjugacy classes,
so for each $\cM$ in the following list, there are also conjugacy classes
$-\cM $ and $\pm \cM^{-1}$.
There are an infinite number of parabolic $SL(2,\Z)
$ conjugacy classes with $Tr({\cal M})=2$, represented by $T^n$:
$$
{\cal M}_{I_n}=\pmatrix{ 1 & n \cr 0 & 1 \cr }
\eqn\mondp$$
for integer $n$. 
 There are three elliptic
$SL(2,\Z)
$ conjugacy classes with $Tr({\cal M})<2$, represented by
$$ {\cal M}_{II}=\pmatrix{ 1 & 1 \cr -1 & 0 \cr },  \qq {\cal 
M}_{III}=\pmatrix{ 0
&  1 \cr -   1 & 0 \cr } , \qq  {\cal M}_{IV}=\pmatrix{ 0 & 1 \cr -1 & -1 \cr 
}
\eqn\monde$$  
There are   an infinite number of hyperbolic $SL(2,\Z)
$ conjugacy classes with $Tr({\cal M})>2$, represented by
$${\cal M}_{H_n}=\pmatrix{ n & 1 \cr -1 & 0 \cr }, 
\eqn\mondh$$  
for integers $n$ with  $|n| \ge 3$, together with
sporadic monodromies ${\cal M}_t$ of trace $t$.
$$\eqalign{
{\cal M}_{8}&=\pmatrix{ 1 & 2 \cr 3 & 7 \cr },  \qq {\cal 
M}_{10}=\pmatrix{ 1
&  4 \cr 2 & 9 \cr } , \qq  {\cal M}_{12}=\pmatrix{ 1 & 2 \cr 5 & 11 \cr 
}
\cr 
{\cal M}_{13}&=\pmatrix{ 2 & 3 \cr 7 & 11 \cr 
},\qq {\cal M}_{14}=\pmatrix{ 1 & 2 \cr 6 & 13 \cr 
}, \dots
\cr}
\eqn\monds$$  
where this is the complete list of sporadic classes for
$3\le t\le 15$ [\DeWolfePR].

For any $\cal M$,
the monodromies $\cal M$ and ${\cal M}^{-1}$ define physically equivalent theories, related
by changing the mass parameter $m \to -m$.
The $SL(2,\Z)$ element $-\II$ acts trivially on the scalars,
 and   reverses the sign of the 2-form potentials.
 The relation between the theory obtained  by a
Scherk-Schwarz reduction with monodromy
${\cal M}$ and that with monodromy
$-{\cal M}$ 
will be discussed in [\atish].

Then the physically distinct  theories obtained by Scherk-Schwarz reduction of 
the
IIB string theory are those with monodromies  \mondp,\mondh, consisting of two infinite 
series, each
labelled by  an integer $n$, the three exceptional cases with monodromies \monde, and
the hyperbolic monodromies \monds. 
(The ones corresponding to \mondp,\monde,\mondh\ are also discussed in [\bergnew].)
These will now be compared
to the three classes of classical supergravity theories obtained with 
monodromies
\monc. The theories given by reduction with monodromy
  ${\cal M}_{I_n}$ correspond to gauging the parabolic subgroup of $SL(2,\R)$
generated by
$M_p(a)$ in \gens. This reduction was first performed in [\bergy], and was shown 
to
give the same 9-dimensional supergravity   as the conventional (untwisted) 
reduction
of the massive IIA supergravity theory of Romans  to 9 dimensions, with mass
parameter
$m=a$. In the quantum theory, this mass is quantized [\bergy], $a=n$ for some 
integer
$n$, as was   seen from a different point of view in [\GHT]. The 
monodromies ${\cal M}_{H_n}$, $|n| \ge 3$ and $\cM_t$ given in \monds\ are conjugate to the
hyperbolic 
gauging
with monodromy
$\cM_h(a)$ and will  again  give a quantum-consistent theory.

 The elliptic monodromy $\cM_e (\th)$ gives the gauging of the compact
$SO(2)$ subgroup of $SL(2,\R)$,  giving the  $SO(2)$-gauged supergravity 
discussed
in [\cow] with mass parameter $m=\th$. In the quantum theory, the monodromies 
$\cM_{II},\cM_{III},\cM_{IV}$
  arise from $SO(2)$ gaugings at special values of the angle $\th$. The 
monodromy
$\cM_{III}$ is clearly  a rotation through $\pi/2$,
 $$\cM_{III}=\cM_e(\pi/2)
\eqn\abc$$
 while the other two are $SL(2,\R)$ conjugate to rotations through $\pi/3, 
2\pi/3$,
i.e. there are matrices $U,V$ in $SL(2,\R)$ such that
$$\cM_{II}=U\cM_e(\pi/3)U^{-1} ,\qq
\cM_{IV}=V\cM_e(2\pi/3)V^{-1}
\eqn\abc$$ Thus in the Scherk-Schwarz reduction with elliptic monodromy giving 
an $SO(2)$
gauging, quantum consistency forces the \lq mass parameter' $\th$ to take the
discrete values $n\pi/2$ or $n\pi/3$ for integer $n$, giving just three  
non-trivial
physically distinct cases, $\th =\pi/2,\pi/3,2\pi/3$.

Thus  Scherk-Schwarz dimensional reduction leads to a quantization of the mass
parameters in the 9-dimensional gauged supergravity theories. The Scherk-Schwarz
reduction of IIB string theory on a circle with monodromy
$\cM\in SL(2,\Z)$  can be viewed as F-theory reduced on a $T^2$ bundle over a 
circle
with monodromy 
$\cM$ [\mf], and  monodromies in the same $SL(2,\Z)$ conjugacy class give 
equivalent
bundles, so the quantization condition has a topological origin. The M-theory 
dual
of these reductions was given in [\mf].

BPS solutions of these 9-dimensional theories have been considered in 
[\bergy,\cow].
The parabolic gauged theory has an exponential potential, so that
the only critical point is for   infinite values of the scalars. It has no
maximally supersymmetric vacuum, but has half-supersymmetric domain wall
solutions in which the wall separates regions  with
different values of the quantized mass $m$ [\bergy]. The theories on either side 
of
the wall, with masses $m,m'$, are related by an $SL(2,\Z)$ transformation 
$\cM_{I_n}$
with parameter $n=m-m'$. The $SO(2)$ gauged theory has a 
potential with a minimum at which the potential vanishes.
It thus has a
Minkowski vacuum [\cow], which breaks all supersymmetries [\mee].
It also has   BPS
domain wall solutions that separate regions  with mass parameters
$\th=\pm \th _0$ [\cow]. For
$\th_0=\pi/2$, the theory with mass parameter $-\pi/2$ is obtained from
that with mass parameter $\pi/2$ by acting  with
the $SL(2,\Z)$ transformation $-\II$, which acts by changing the sign of the 
2-form
gauge fields. The theory with $\th =-\pi/3$ is obtained from  the one with  
 $\th =2\pi/3$ by acting with  the $SL(2,\Z)$ transformation $-\II$. The 
theories
with mass parameters $\th =\pm \pi/3$ are related by an $SL(2,\Z)$ 
transformation
$\cM_{IV}$, conjugate to a rotation through $2\pi/3$.
General BPS domain walls of all three classes of supergravity theories with
monodromies \monc\ will be considered in [\bergnew], together with further
properties of these theories, and in particular the structure of the scalar
potentials.

\refout
\bye